\documentclass[aps,jcp,superscriptaddress,amsmath,amssymb]{revtex4-1}

\usepackage{comment}
\usepackage{tikz}
\usetikzlibrary{trees}
\usetikzlibrary{decorations.pathmorphing}
\usetikzlibrary{decorations.markings}
\tikzset{
    photon/.style={decorate, decoration={snake,segment length=1.5mm}, draw=black},
    coulomb/.style={dotted},
    electron/.style={draw=black, postaction={decorate},
        decoration={markings,mark=at position .55 with {\arrow[draw=black]{>}}}}, 
    gluon/.style={decorate, draw=magenta,
        decoration={coil,amplitude=4pt, segment length=5pt}},
    boundelectron/.style={thick, double},
    transverse/.style={dashed}
}

\usepackage{graphicx}
\usepackage{dcolumn}
\usepackage{bm}
\usepackage{rotating}

\newcolumntype{.}{D{.}{.}{8}}

\usepackage[utf8]{inputenc}
\usepackage{physics}
\usepackage{amsmath, amssymb}
\usepackage{tabularx,booktabs,array,dcolumn}
\usepackage{setspace}
\usepackage{vmargin}
\usepackage{xcolor}
\usepackage{graphicx,psfrag,subfigure}

\usepackage{float}
\usepackage[all]{xy}

\usepackage{bm}
\usepackage{mathtools}
\usepackage{physics}
\usepackage[english]{babel}

\newcommand{\bos}[1]{\boldsymbol{#1}}
\newcommand{\mr}[1]{\mathrm{#1}}

\def\Eh{E_\text{h}}

\def\iim{\mr{i}}

\def\bPsi{\bos{\Psi}}

\def\bPhi{\bos{\Phi}}
\def\np{n}
\def\nnuc{N_\text{nuc}} 

\def\nbas{{N_\text{b}}}
\def\nbasnull{{N_\text{b}^{(0)}}}
\def\nbasone{{N_\text{b}^{(1)}}}

\def\nnuc{{N_\text{nuc}}}
\def\deltaexp{\left\langle\delta\right\rangle}

\def\bp{\bos{p}}
\def\br{\bos{r}}
\def\bs{\bos{s}}

\def\bA{\bos{A}}

\def\bPsi{\bos{\Psi}}

\def\som{Supplementary Material}

\def\Hetwop{He$_2^+$}
\def\Htwo{{H$_2$}}
\def\Hthreeplus{H$_3^+$}
\def\Hthree{H$_3$}
\def\Sgp{\Sigma_\text{g}^+}
\def\Sup{\Sigma_\text{u}^+}
\def\Pig{\Pi_\text{g}}
\def\Piu{\Pi_\text{u}}

\usepackage[unicode]{hyperref}
\usepackage{soul}

\def\blog{\text{ln}k_0}

\usepackage[unicode]{hyperref}
\hypersetup{
   unicode=true,          
   plainpages=false,
   colorlinks=true,       
   linkcolor=red,          
   linkcolor=blue,          
   citecolor=blue,        
   urlcolor=blue           
}

\usepackage{url}

\begin{document}

\title{%
Evaluation of the Bethe logarithm: from atom to chemical reaction
}

\author{D\'avid Ferenc} 
\author{Edit M\'atyus} 
\email{edit.matyus@ttk.elte.hu}
\affiliation{ELTE, Eötvös Loránd University, Institute of Chemistry, 
Pázmány Péter sétány 1/A, Budapest, H-1117, Hungary}

\date{\today}

\begin{abstract}
\noindent %
A general computational scheme for the (non-relativistic) Bethe logarithm is developed opening the route to `routine' evaluation of the leading-order quantum electrodynamics correction (QED) relevant for spectroscopic applications for small polyatomic and polyelectronic molecular systems. 
The implementation relies on the Schwartz method and minimization of a Hylleraas functional. 
In relation with electronically excited states, a projection technique is considered, which ensures positive definiteness of the functional over the entire parameter (photon momentum) range.
Using this implementation, the Bethe logarithm is converged to a relative precision better than 1:10$^3$ for selected electronic states of the two-electron H$_2$ and H$_3^+$, and the three-electron He$_2^+$ and H+H$_2$ molecular systems. 
The present work focuses on nuclear configurations near the local minimum of the potential energy surface, but the computations can be repeated also for other structures. 
\end{abstract}

\maketitle
\clearpage

\noindent %
\section{Introduction}
\noindent %
Relativistic and quantum electrodynamics (QED) corrections represent important contributions for the theoretical, high-precision description of few-particle atomic- and molecular systems. 
For few-electron atoms and molecules with low nuclear charge numbers, the leading-order relativistic correction is routinely evaluated as the expectation value of the Breit--Pauli Hamiltonian \cite{Breit1929,BrRa51,Sa52,BeSaBook57,DyallFaegriBook07}.
At the same time, pinpointing the value of the corrections to high precision is a challenging task due to the singular behavior of the correction operators. Special care must be taken when using approximate wave functions, which fail to exactly describe the electron-nucleus and electron-electron coalescence points \cite{PaCeKo05,JeIrFeMa22}. 

The computation of the leading-order QED terms is more involved, mostly due to the evaluation of the logarithm corresponding to the `mean excitation energy' introduced by Hans Bethe in 1947 \cite{bethe1947}. The term is commonly referred to as the (non-relativistic) `Bethe logarithm', $\blog$, and it was used to account for most of the `Lamb shift', the discrepancy between Dirac's theory of the hydrogen atom and the experimental observation \cite{LaRe47}. The leading-order QED correction terms have been later generalized for few-particle systems beyond the hydrogen atom \cite{araki57,sucherPhD1958,BrLa68}. 

For the two-electron helium atom, the Bethe logarithm was first computed precisely by Schwartz in 1961 \cite{schwartz1961}, and it had remained the most accurate value for decades. 
For helium-like ions with a higher $Z$ nuclear charge number, a formal expression based on the $1/Z$ series was presented by Goldman and Drake \cite{Goldman1983}, and later, explicit computations have been carried out \cite{DrGo99,BhDr98}. 
In 2019, the Bethe logarithm was reported by Korobov to 12--14 significant digits for several states of the helium atom with the inclusion of also the nucleus in the (non-relativistic) quantum system \cite{Korobov2019}.

For the three-electron lithium and the four-electron beryllium atoms, the Bethe logarithm was reported with a fixed nucleus, and later, by including also the nucleus in the quantum system 
\cite{yan_bethe_2003,PaKo03,pachucki_2006,stanke_2009,PuKuPa2013}.

The H$_2^+$ ion was the first molecular system, for which the Bethe logarithm was evaluated at a series of fixed nuclear configurations in 1992 \cite{BuJeMoKo92}, and later, by treating H$_2^+$ and HD$^+$ as a three-particle system \cite{Korobov2004,Korobov2006,Korobov2012,Korobov2012-velocity}. 
Regarding the two-electron H$_2$ molecule, the Bethe logarithm was reported in 2009 for the ground electronic state at a series of fixed nuclear configurations \cite{PiLaPrKoPaJe09}. In 2019, the four-particle Bethe logarithm was determined for the rovibronic ground state of \Htwo{} treated as a four-particle quantum system \cite{PuKoCzPa2019}. 

The precise evaluation of the Bethe logarithm is a computationally and technically demanding task, and for this reason, the leading-order QED correction is often only estimated or approximated, \emph{e.g.,} by  exploiting the weak dependence of $\blog$ on the number of electrons \cite{FeKoMa2020}. At the same time, it is difficult to rigorously assess the uncertainty arising from the approximations.  
For this reason, it is necessary to extend the range of (molecular) systems for which the precise $\blog$ value is known.

In this work, we describe the implementation of the Schwartz method \cite{schwartz1961} in the QUANTEN computer program  aiming at general applicability for small polyatomic and polyelectronic systems. 
QUANTEN (QUANTum mechanical treatment of Electrons and atomic Nuclei) is an in-house developed, general-purpose molecular physics platform for theoretical developments in precision physics and spectroscopy, and includes by now computation of
single- and multi-state non-adiabatic corrections to the non-relativistic energy \cite{FeMa2019HH,MaFe22}, pre-Born--Oppenheimer energy \cite{FeMa2019EF}, energy lower bounds \cite{IrJeMaMaRoPo22}, perturbative relativistic and QED corrections \cite{FeKoMa2020,JeIrFeMa22}, and variational relativistic energies \cite{JeFeMa21,JeFeMa22,FeJeMa22,FeJeMa22b}.

The structure of the present paper is as follows. The first section defines the methodological background, including aspects of point-group symmetry and projection, then we collect the technical and computational details, finally we report the Bethe logarithm values obtained during this work.

\section{Methodology \label{sec:method}}
We start with the solution of the Schrödinger equation for $\np$ electrons and $N_\text{nuc}$ clamped nuclei using the non-relativistic Hamiltonian (in Hartree atomic units) for the $\psi^{(0)}$ reference state 
\begin{gather}
    H\psi^{(0)}=E\psi^{(0)} \label{eq:Scrödinger} \\
    H
    =
    \frac{1}{2}\sum_{i=1}^\np p_i^2 -\sum_{i=1}^\np \sum_{a=1}^{N_\text{nuc}} \frac{Z_a}{r_{ia}} +\sum_{i=1}^\np \sum_{j>i}^\np \frac{1}{r_{ij}} + \sum_{a=1}^{N_\text{nuc}}
     \sum_{b>a}^{N_\text{nuc}} \frac{Z_aZ_b}{r_{ab}} \; .
    \label{eq:H0}
\end{gather}

The leading-order ($\alpha^2\Eh$ with the $\alpha$ fine structure constant) `relativistic' correction to the non-relativistic energy was first obtained by Breit using the Pauli approximation \cite{Breit1929,BeSaBook57,DyallFaegriBook07}.

The leading-order ($\alpha^3\Eh$) QED correction \cite{araki57,sucherPhD1958,BrLa68} written in a form often used in applications, \emph{e.g.,} Ref.~\citenum{FeKoMa2020}, is
\begin{align}
    E^{(3)} 
    &= 
    \alpha^3 \frac{4}{3} 
    \sum_{i=1}^\np 
      \left( \ln \frac{1}{\alpha^2} - \blog  +\frac{19}{30} \right) \sum_{a=1}^\nnuc{} Z_a \langle \delta(\br_{ia}) \rangle \nonumber \\
    &+\alpha^3 
    \sum_{i=1}^\np{} \sum_{j>i}^\np{} 
      \left[ \left( \frac{14}{3}\ln \alpha + \frac{164}{15}\right) \langle\delta(\br_{ij})\rangle -\frac{14}{3}Q \right] \; ,
\end{align}
where $\langle \delta(\bos{r}_{ia}) \rangle$ and $\langle \delta(\bos{r}_{ij}) \rangle$ label expectation values with the non-relativistic wave function, and
$Q$ is the Araki--Sucher term for the same non-relativistic state \cite{araki57,sucherPhD1958,Sucher58}. 

In what follows, all corrections are written for the $(E_0,\psi_0^{(0)})$ non-relativistic ground-state energy and wave function, but similar expressions apply for $(E_n,\psi_n^{(0)})$ excited states.
The quantities $\langle \delta(\bos{r}_{ia}) \rangle$, $\langle \delta(\bos{r}_{ij}) \rangle$, and $Q$ can be efficiently evaluated using the integral transformation technique \cite{PaCeKo05}. 

The Bethe logarithm, $\blog$, for 
$(E_0,\psi_0^{(0)})$ 
is written as  \cite{PaKo03}
\begin{align}
    \blog
    &= 
    -\frac{1}{2\pi \langle \delta \rangle} 
    \matrixel*{\psi^{(0)}_0}{\grad(H-E_0)\ln|2(H-E_0)/\Eh{}|\grad}{\psi^{(0)}_0} \, , \label{eq:BL0}
\end{align}
where the following short-hand notation is defined for later convenience,
\begin{align}
  \delta 
  =
  \sum_{i=1}^\np \sum_{a=1}^{\nnuc} Z_a \delta(\br_{ia}) 
  \quad\ 
  \text{and}
  \quad\ 
  \grad 
  = 
  \sum_{i=1}^{\np} \grad_i 
\end{align}
with $\nabla_\alpha =\sum_{i=1}^{\np} \nabla_{i\alpha}$ and the $\alpha=x(1),y(2),z(3)$ Cartesian degrees of freedom.
(We also note that $1\ \text{Ry}=1/2 \Eh$ commonly used in the older literature.) 
Furthermore, 
\begin{align}
  \left\langle \delta \right\rangle 
  &= 
  \matrixel*{\psi^{(0)}_0}{ \delta }{\psi^{(0)}_0}
  \quad\ 
  \text{and}
  \quad\  
  \left\langle \grad^2 \right\rangle 
  = 
  \matrixel*{\psi^{(0)}_0}{ \grad^2 }{\psi^{(0)}_0}  \; .
\end{align}

Following Schwartz \cite{schwartz1961} and Korobov \cite{Korobov2019}, we do not use Eq.~\eqref{eq:BL0} directly, but turn to the unrenormalized integral representation and explicitly subtract the $-\langle \grad^2\rangle \Lambda / (2\pi \langle \delta \rangle)$ mass-renormalization counterterm from the integral over the photon momenta,
\begin{align}
  \blog 
  =
  \frac{1}{2\pi \deltaexp}
  \lim_{\Lambda\rightarrow\infty}  
  \left[%
    \int_0^\Lambda k J(k)\  \dd k
    +
    \langle \grad^2 \rangle \Lambda
    + 
    2\pi \deltaexp \ln \Lambda
  \right]  
  +\text{ln}2
  \; 
  \label{eq:BL1}
\end{align}
with
\begin{align}
   J(k)
   =
   -
   \langle \psi_0^{(0)} | \grad 
   (H-E_0+k)^{-1}
   \grad | \psi_0^{(0)} \rangle
   = 
   \sum_{j\neq 0}
     \frac{%
       |\langle \psi_0^{(0)} | \grad \psi_j^{(0)}\rangle |^2
     }{%
       E_j-E_0+k
     } 
    \label{eq:Jsos}  \; ,
\end{align} 
where the summation includes also integration over the continuum states of the reference problem, Eqs.~(\ref{eq:Scrödinger})--(\ref{eq:H0}). 
In Eqs.~(\ref{eq:BL1})--(\ref{eq:Jsos}) and in the following computations, all energies are understood in $\Eh$ units.
We use Eqs.~(\ref{eq:BL1})--(\ref{eq:Jsos}) for further computations, and the following relations are useful to understand the connection with the more common form Eq.~(\ref{eq:BL0}).
The $\ln 2$ term in Eq.~(\ref{eq:BL1}) can be obtained from Eq.~(\ref{eq:BL0}) using the following relationship (pp. 99 of Ref.~\citenum{BeSaBook57})
\begin{align}
  \matrixel{\psi^{(0)}_0}{\grad (E_0-H) \grad}{\psi^{(0)}_0} 
  &=
  \matrixel{\psi^{(0)}_0}{\left[H,\grad\right]\cdot\grad}{\psi^{(0)}_0}
  \nonumber \\
  &=
  \matrixel{\psi^{(0)}_0}{\grad\cdot\left[\grad,H\right]}{\psi^{(0)}_0} \nonumber \\
  &=
  \frac{1}{2}\matrixel{\psi^{(0)}_0}{\left[\left[H,\grad\right],\grad\right]}{\psi^{(0)}_0} 
  \nonumber \\
  &= 
  2\pi \sum_{i=1}^\np \sum_{a=1}^\nnuc Z_a \matrixel{\psi^{(0)}_0}{\delta(\br_{ia})}{\psi^{(0)}_0} 
  =
  2\pi \langle \delta\rangle  \; .
\end{align}
Furthermore, the connection of Eq.~(\ref{eq:BL1}) and Eq.~(\ref{eq:BL0}) can be seen better by merging  Eqs.~(\ref{eq:BL1}) and (\ref{eq:Jsos}) as
\begin{align}
  \blog 
  =
  -\frac{1}{2\pi \deltaexp}
  \langle \psi_0^{(0)} | \grad 
  \lim_{\Lambda\rightarrow\infty}  
  \left[%
    \int_0^\Lambda 
    \left( \frac{k}{H-E_0+k} -1 \right)  \dd k
    + 
    (H-E_0) \ln \Lambda
  \right] \grad 
  | \psi_0^{(0)} \rangle
  +\text{ln}2 \; ,
\end{align}
and using the elementary integral (with $a=H-E_0$)
\begin{align}
  &\lim_{\Lambda\rightarrow\infty}
     \left[
       \int_0^\Lambda \left( \frac{k}{a+k} - 1 \right) \dd k  + a \ln \Lambda
     \right] 
  =
  \lim_{\Lambda\rightarrow\infty}
     \left[
       \int_0^\Lambda \left(-\frac{a}{a+k}\right) \dd k  + a \ln \Lambda 
     \right] \nonumber \\
  =
  &\lim_{\Lambda\rightarrow\infty}
     \left[
       -a(\ln|a+\Lambda| - \ln|a|) + a \ln \Lambda 
     \right] 
  = a \ln |a| +
     \lim_{\Lambda\rightarrow\infty}
       a \ln \Bigg|\frac{\Lambda }{a+\Lambda}\Bigg|
  = a \ln |a|    \; .
\end{align}

Although $J(k)$ is written in a form of a `sum' over states, Eq.~(\ref{eq:Jsos}), formal analysis shows that contribution of distant states (in energy) is more important than contribution of electronic states near $E_0$.
An efficient evaluation route for $J(k)$ is explained later in this section. First, the integral for the $k$ photon momentum is discussed. In general, this integral can be calculated numerically. For numerical stability in finite precision arithmetic \cite{schwartz1961}, it is convenient to split the $[0,\Lambda]$ interval (with $\Lambda\rightarrow \infty$) into three parts. 

\subsection{Integration with respect to the photon momentum \label{sec:integrate}}
For the low-momentum part, $k\in [0,\kappa_1]$, the counterterm can be explicitly subtracted without any numerical problem. So, we can write,
\begin{align}
  (\blog)_\text{low} 
  &=  
  \frac{1}{2\pi \langle \delta \rangle}
  \left[%
    \int_0^{\kappa_1} k J(k)\ \dd k  + \langle \grad^2 \rangle  \kappa_1 
  \right]
  +  
  \ln \kappa_1  
  \; .
  \label{eq:low}
\end{align}
As the $\kappa_1$ upper limit of the integration increases, the $\int_0^{\kappa_1} k J(k)\ \text{d}k$ integral becomes very large (diverges in the $\kappa \rightarrow \infty$ limit), and explicit subtraction of the mass-renormalization counterterm, as in Eq.~(\ref{eq:low}), would introduce large numerical errors (small difference of two large numbers).

For the intermediate (`middle', mid) range, $k\in[\kappa_1,\kappa_2]$, it is possible to stabilize the result, by performing the operations within the integrand (with $\int k^{-1}\ \dd k= \text{ln} k + c$),
\begin{align}
  (\blog)_\text{mid} 
  &=   
  \frac{1}{2\pi  \langle \delta \rangle} 
  \int_{\kappa_1}^{\kappa_2} 
    \left[ %
      k J(k) + \langle \grad^2 \rangle + 2\pi  \langle\delta\rangle k^{-1} 
    \right] 
    \text{d} k
    \; .
  \label{eq:mid}
\end{align}
As $k$ approaches to infinity in the high momentum range, $k\in[\kappa_2,\Lambda)$ with $\Lambda\rightarrow \infty$, precise numerical computation of $J(k)$ at selected $k$ values (`grid points') becomes problematic, and for this reason, the asymptotic, high-$k$ expansion of the integrand---derived for the hydrogen-like atoms \cite{schwartz1961,Korobov2012-velocity}---is used to arrive at an accurate result, 
\begin{align}
  k > \kappa_2:\quad &
  k J(k) + \langle \grad^2 \rangle + 2\pi  \langle\delta\rangle k^{-1}   \nonumber \\
  &\approx
  \frac{4\pi}{k^2}\sum_{i=1}^\np \sum_{a=1}^{\nnuc} Z_a^2
  \left[%
    (2k)^{\frac{1}{2}}-Z_a \ln k 
  \right]
  \langle\delta ( \br_{ia})\rangle 
     +\frac{1}{k^2}\sum_{m=0}^\mathcal{M} c_m k^{-\frac{m}{2}} = \mathcal{K}(k) \; .
     \label{eq:expansion}
\end{align}
The $c_m$ coefficients are fitted to the middle-range values of $kJ(k)$, and based on some numerical experimentation, we have chosen $\mathcal{M}=5$ for the fit. Using the fitted, analytic expression, the integral over the $(\kappa_2,\infty)$ high-momentum range is evaluated analytically. According to our experience, the results are not sensitive to the precise value of $\kappa_1$ and $\kappa_2$. 

Regarding the $\Lambda \rightarrow \infty$ limit of the high-momentum-range integral,
\begin{align}
  (\blog)_\text{high}
  =
  \lim_{\Lambda\rightarrow\infty} 
  \int_{\kappa_2}^\Lambda \mathcal{K}(k)\ \dd k \; ,
\end{align}
we can observe in Eq.~(\ref{eq:expansion}) that the limit of the $\Lambda$-dependent terms 
go to zero at the order of $1/\sqrt{\Lambda}$, and thus, we obtain a finite limit.
As a result, the Bethe logarithm defined in Eq.~(\ref{eq:BL0}) is obtained as
\begin{align}
  \blog
  =
  (\blog)_\text{low}
  +
  (\blog)_\text{mid}
  +
  (\blog)_\text{high}
  +\ln 2 \; .
\end{align}

\subsection{Evaluation of the sum-over-states expression \label{sec:sos}}
In this procedure, the key quantity, $J(k)$, is written as a sum-over-states expression in Eq.~(\ref{eq:Jsos}), and it must be converged for several $k$ values to arrive at a precise $\ln k_0$ result. 
Direct summation over the non-relativistic eigenstates used as a basis converges slowly with respect to the  basis size (number of eigenstates), and for this reason, the $J(k)$ values are computed differently (using a separate, `auxiliary' basis set optimized for the problem).
The core idea, which allows us to avoid explicit summation over all eigenstates, was introduced by Hylleraas for the evaluation of the perturbative corrections due to the electron-electron Coulomb interaction to the non-relativistic ground-state energy of the helium atom \cite{Hylleraas1930} (for a modern introduction in the context of non-relativistic electronic structure theory, see for instance Ref.~\citenum{mayer_simple_2003}). 
The Hylleraas functional method can be adopted for a variety of sum-over-states (perturbation theory) problems. Most recently, we used it for the computation of the
non-adiabatic mass correction to coupled electronic states \cite{MaFe22}.

For the particular case of $J(k)$ in relation with the non-relativistic Bethe logarithm,
a `so-called' perturbed wave function, $\bPsi^{(1)}=(\Psi^{(1)}_x,\Psi^{(1)}_y,\Psi^{(1)}_z)$, is obtained by variational minimization of the Hylleraas functional,
\begin{align}
  \mathcal{J}_k\left[\bos{\Phi}\right] 
  = 
  \matrixel*{\bos{\Phi}}{H-E_0+k}{\bos{\Phi}} 
  + 
  2
  \langle 
    \bos{\Phi} | 
    \grad \psi^{(0)}_0
  \rangle \; .
  \label{eq:Hyfun}
\end{align}
The $\bPhi=\bPsi^{(1)}$ function, which makes $\mathcal{J}_k$ stationary (minimal) 
is provided by the linear equation solved for $\bos{\Psi}^{(1)}$,
\begin{align}
    (H-E_0+k)\ket*{\bos{\Psi}^{(1)}} + \ket*{\grad\psi^{(0)}_0} =0 \; .
    \label{eq:lineq1} 
\end{align}
If it is solved, then $J(k)$ is obtained as
\begin{align}
    J(k)
    =
    \langle \psi^{(0)}_0 | \grad \bPsi^{(1)} \rangle
    =
    \sum_\alpha
    \langle \psi^{(0)}_0 | \nabla_\alpha \psi^{(1)}_\alpha \rangle
    \; .
    \label{eq:Jk}
\end{align}
The three Cartesian components of the perturbed wave function, $\Psi^{(1)}_\alpha$ with $\alpha=x(1),y(2),z(3)$,  are expanded over a set of auxiliary $\varphi^{(1)}_{\alpha i}$ basis functions (Computational details section), 
\begin{align}
  \psi^{(1)}_\alpha
  =
  \sum_{i=1}^\nbasone{} 
    c_{i\alpha}^{(1)}\, 
    \varphi^{(1)}_{\alpha i} \; .
    \label{eq:basis}
\end{align}
The superscript `(1)' emphasises that the basis functions are generated and optimized to minimize the Hylleraas functional, Eq.~(\ref{eq:Hyfun}), and they are distinct from that of the basis set used for the `(0)' reference wave function, Eq.~(\ref{eq:H0}).
This setup makes it possible to systematically (variationally) improve $J(k)$, through Eq.~(\ref{eq:Jk}), by systematic improvement 
of the basis representation (and parameterization) of $\bPsi^{(1)}$, by mimimization of the Hylleraas functional, Eq.~(\ref{eq:Hyfun}),
instead of the energy functional.

For a given basis set and parameterization, we substitute the Eq.~(\ref{eq:basis}) expansion into Eq.~\eqref{eq:lineq1}, and multiply from the left by $\bra*{\varphi_{\alpha j}^{(1)}}$ ($j=1,\ldots,\nbasone$),
\begin{align}
  \sum_{i=1}^{\nbasone} 
    c_{\alpha i}^{(1)} \matrixel*{\varphi_{\alpha j}^{(1)}}{H-E_0+k}{\varphi_{\alpha i}^{(1)}} 
  + 
  \sum_{k=1}^{\nbasnull}
    c_k^{(0)}\bra*{\varphi_{\alpha j}^{(1)}}\nabla_\alpha \varphi^{(0)}_k \rangle =0 \; .
\end{align}
The $c_{\alpha i}^{(1)}$ linear combination coefficients are obtained as the solution of the linear equation (by standard linear algebra routines), 
\begin{align}
     \bos{M}_\alpha \bos{c}^{(1)}_\alpha &= -\bos{b}_\alpha \; 
     \label{eq:lineq2}
\end{align}
with
\begin{align}
  (M_\alpha)_{ij} 
  &= 
  \matrixel*{\varphi_{\alpha i}^{(1)}}{H-E_0+k}{\varphi_{\alpha j}^{(1)}}
\end{align}
and
\begin{align}
  (b_{\alpha})_j
  &= 
  \sum_{k=1}^{\nbasnull} c_{k}^{(0)}\bra*{\varphi_{\alpha j}^{(1)}}\nabla_\alpha
  \varphi^{(0)}_k \rangle  \; .
\end{align}

The selection (and parameter optimization) of the $\varphi_{\alpha j}^{(1)}$ basis functions is carried out
based on the minimization condition of the Hylleraas functional, Eq.~(\ref{eq:Hyfun}). Further technical details are explained in the Computational details section.

Finally, we note that the value of $J(0)$ is exactly known based on the Thomas--Reiche--Kuhn sum rule (summarized in the \som),
\begin{align}
  J(0)=\frac{3}{2} \np \; .
\end{align}
Comparison of this exact result and the numerically computed $J(0)$ value 
is used to estimate the numerical uncertainty \cite{pachucki_bethe_2003} of the computed $J(k)\ (k>0)$ values (for which no exact results are known).

\subsection{Bethe logarithm for electronically excited states \label{sec:exstate}}
By replacing the ground-state energy and wave function, $(E_0,\psi^{(0)}_0)$, in the previous expressions with the same quantities $(E_i,\psi^{(0)}_i)$ of some excited state can be considered as a starting point for computing the Bethe logarithm for excited states.
The computation of the perturbed wave function by minimization of the corresponding Hylleraas functional, Eq.~(\ref{eq:Hyfun}), is viable, if the functional is bounded from below. 
Unfortunately, for an electronically excited state,
there are $k$ values, $k<E_i-E_n$ ($n=0,\ldots$), 
for which the lower-boundedness condition is not fulfilled
(the perturbed wave function component overlaps with the lower-energy state wave function, unless some special point-group symmetry relation applies). 

Then, we need to consider the minimization of Eq.~(\ref{eq:Hyfun}) with the auxiliary orthogonality condition of the $\Psi^{(1)}_\alpha$ perturbed wave function components to the lower-lying states (see also Ref.~\citenum{MaFe22}), $(E_n,\psi_n)$ with $E_n<E_i-k$ (we collect the relevant indexes in set $\mathcal{L}$),   
\begin{align}
  \matrixel{\bPhi}{P}{\bPhi}=0 && P = \sum_{n\in\mathcal{L}} \dyad{\psi_n} \; , \label{eq:orthog}
\end{align}
which leads to the equation for $\bPhi=\bPsi^{(1)}$
\begin{align}
    (H-E_i+k)P^\perp|\bPsi^{(1)}\rangle+P^\perp |\grad \psi_i\rangle =0 \; ,
    \quad \text{with}\quad P^\perp = 1-P
    \; .
\end{align}
These projected equations (in practice, the projector is constructed in the actual auxiliary basis set) can be used to select and optimize the basis set in a variationally stable fashion. 
Then, using the optimized basis set, the original, unprojected equation, Eq.~(\ref{eq:lineq1}), is solved to obtain the final $J(k)$ value.

\subsection{Symmetry considerations\label{sec:symm}}
In the present work, 
we consider 
the $D_{\infty \text{h}}$ group (\Htwo{} and \Hetwop{}, equilibrium structure, the nuclei fixed  along the $z$ axis symmetrically around the origin), 
the $D_{3 \text{h}}$ group (\Hthreeplus{}, equilibrium structure, the nuclei fixed in the $x-y$ plane, symmetrically around the origin), 
and $C_{\infty \text{v}}$ (\Hthree{}, equilibrium structure, the nuclei fixed along the $z$ axis, and the nuclear center of mass fixed at the origin). Furthermore, the $O(3)$ group is used for atomic states (He with the nucleus centered at the origin).

If the $\psi^{(0)}$ reference state is totally symmetric, 
then the symmetry of the three components of the perturbed wave function correspond to the symmetry of the $x$, $y$, and $z$ coordinates in the corresponding point group (Table~\ref{tab:sym}).
For $O(3)$ (He atom), all three components ($\alpha=x,y,z$) have the same symmetry. 
For the rest of the systems considered in this work, the $x$ and $y$ components belong to the same irreducible representation (irrep), and the $z$ component is distinct. 
In general, for a lower-order group, \emph{e.g.,} $C_{2\text{v}}$, all three $\alpha$ components belong to different irreps.

\begin{table}[h]
    \centering
    \caption{%
      Symmetry species of the reference state and the perturbed wave functions within the groups used to compute the $\blog$ values in Table~\ref{tab:BL}. 
    \label{tab:sym}}    
    \begin{tabular}{@{}c| c @{\ } c @{\ } c @{\ } c @{}}
      \hline\hline\\[-0.35cm]
      Point group   & $\Gamma (\psi^{(0)})$  & $\Gamma (\psi^{(1)}_{x})$ & $(\psi^{(1)}_{y})$ & $\Gamma (\psi^{(1)}_{z})$ \\
      \hline\\[-0.35cm]
      $O(3)$   & $S$ & $P$ & $P$  & $P$ \\
      \hline\\[-0.35cm]
      $D_{\infty \text{h}}$   & $\Sgp$ & $\Piu$ & $\Piu$ & $\Sup$ \\
                       & $\Sup$ & $\Pig$ & $\Pig$ & $\Sgp$ \\
      \hline\\[-0.35cm]
     $C_{\infty \text{v}}$
      & $\Sigma^+$   & $\Pi$   & $\Pi$  & $\Sigma^+$ \\
      \hline\\[-0.35cm]
      $D_{3\text{h}}$         & $A_1'$       & $E'$    & $E'$    & $A_2''$ \\
      \hline\hline
    \end{tabular}
\end{table}

\section{Computational details \label{sec:compdet}}

The methodology has been implemented in the QUANTEN computer program \cite{Matyus2019,FeMa2019HH,FeMa2019EF,FeKoMa2020,JeIrFeMa22,IrJeMaMaRoPo22,JeFeMa21}. 
First, the non-relativistic Schrödinger equation, Eq.~\eqref{eq:Scrödinger}, is solved using a linear combination of antisymmetrized ($\mathcal{A}$), symmetry-adapted $(\mathcal{P}_{\Gamma})$ products of spatial and spin functions,
\begin{align}
  \psi
  = 
  \sum_{i=1}^{\nbas} 
    c_i \varphi_i(\br;\bA_i,\bs_i) 
  = 
  \sum_{i=1}^{\nbas} 
    c_i \mathcal{A} \mathcal{P}_{\Gamma} \phi_i(\br;\bA_i,\bs_i) \; ,
\end{align}
where $\phi_i(\br;\bA_i,\bs_i)$ is written as a product of spatial and spin-functions. The spin functions corresponding to the spin states of the singlet and doublet spin states of the two- and three-electron systems computed in this work are defined in Ref.~\citenum{MaRe12}.

The spatial functions are (floating) explicitly correlated Gaussian functions (ECGs), 
\begin{align}
  f_i(\br;\bA_i,\bs_i) = \exp\left[-(\br-\bs_i)^T (\bA_i\otimes \bos{I}_3) (\br-\bs_i) \right] \; ,
\end{align}
where 
$\bos{I}_3$ is the three-by-three unit matrix and
$\bA_i\in\mathbb{R}^{\np\times \np}$ is a positive-definite, symmetric matrix with elements optimized variationally. Both the $\bA_i$ and $\bs_i$ parameters are optimized variationally by stochastic and deterministic parameter selection \cite{SuVaBook98} and repeated refinement cycles using the Powell algorithm \cite{Po2004}. 
Regarding the $\bs_i \in \mathbb{R}^{3\np}$ vector, during the course of generation and optimization, some simple `constraints' were imposed to facilitate symmetry adaptation.
For an atomic (He) $S$ state, the $\bs$ vector was set to zero, while for a $P$ ($z$) state, it was constrained to the $z$ axis.
For the $\Sigma$ state of the linear systems (H$_2^+$, He$_2^+$, H$_3$),
the $s_x$ and $s_y$ components were set to zero and the $s_z$ component was included in the variational optimization.
For the $A_1'$ and $A_2''$ symmetry states of H$_3^+$, $s_z$ was set to zero and $s_x$ and $s_y$ were optimized variationally. For the $E'$ symmetry state of H$_3^+$, no constraints were imposed and all the $s_x,s_y,$ and $s_z$ elements were optimized variationally.
To adapt the symmetry of the basis functions to the $\Gamma$ irrep of the relevant group, we directly considered the action of the $\mathcal{P}_{\Gamma}$ projector on the `primitive' basis functions (see for example Ref.~\citenum{MaRe12} and \citenum{JeFeMa22}).

There are separate parameter sets optimized for the reference, `(0)', and for the $\alpha=x,y,$ and $z$ perturbed, `(1)', components for several $k$ values. In the present work, we used the series of $k_n=10^n$ with $n=0,1,\ldots,5$, for which  a perturbed basis set was generated by minimizing the Hylleraas functional. 

For this series of $k_n$ values, separate basis sets were generated and optimized. Then, these basis sets were merged and the value of $J$ was evaluated over the $\xi_j$ quadrature points ($j=0,1,\ldots,N_\text{GL}$) in the merged basis set. We used a separate set of Gauss--Legendre (GL) quadrature points for the low- and for the mid-photon-momentum range, $(\blog)_\text{low}$ and $(\blog)_\text{mid}$, including $N_\text{GL}=20-30$ points, to numerically evaluate the integrals in Eqs.~(\ref{eq:low}) and (\ref{eq:mid}). 
The $k$ variable in the $[0,\kappa_1]$ and $[\kappa_1,\kappa_2]$ intervals was scaled to the $[0,1]$ interval of the GL grid.

The resulting $\blog$ value can be considered to be (practically) variational \cite{Korobov2019} due to the Hylleraas functional formalism, with the only numerical deviations arising from the numerical integration and the numerical uncertainty of the auxiliary quantities ($\langle\delta(\br_{ia})\rangle$ and $\langle \bos{\nabla}^2\rangle$ collected in Table~S1 of the \som).

\begin{table}[h]
  \caption{%
    Numerical tests of the reported implementation for He, H$_2^+$, and H$_2$ with respect to benchmark $\blog$ values available in the literature. For the case of H$_2$, the authors of Ref.~\citenum{PiLaPrKoPaJe09} estimated the error to be in the last digit. 
    \label{tab:BLtest}}
 \begin{tabular}{@{}l lc l @{}}
    \hline\hline\\[-0.35cm]
    & & $R$ & \multicolumn{1}{c}{$\ln k_0$} \\
    \cline{1-4} \\[-0.3cm]
    He & 1 $^1S$  &              & $ 4.370\ 160\ 223\ 0703(3)$~\cite{Korobov2019} \\
       &          &              & $4.370\ 154(8) $ [this work] \\
    \cline{2-4} \\[-0.3cm]
       & 2 $^1S$  &                & $4.366\ 412\ 726\ 417(1)$~\cite{Korobov2019} \\
       &          &                & $4.366\ 41(1)$  [this work] \\
    \cline{1-4} \\[-0.3cm]
    H$_2^+$ & $\tilde{X}\ ^2\Sgp$ & $2.00 $ & $3.012\ 508\ 830$~\cite{korobov_calculation_2013} \\
            &                     &         & $3.012\ 52(1)$  [this work] \\
    \cline{1-4} \\[-0.3cm]
    H$_2$ &	$ X\ ^1\Sgp$ & $1.40 $ & $3.018\ 55(1)$~\cite{PiLaPrKoPaJe09}  \\
          &              &         & $3.018\ 55(3)$  [this work] \\
    \hline\hline
  \end{tabular}
\end{table}

\section{Numerical results \label{sec:numres}}
First, as a test of our implementation, the Bethe logarithm was calculated for the helium atom in its $1\ ^1S$ ground and $2\ ^2S$ excited state, and for the ground state of the \Htwo{}$^+$ molecular ion with $R=2.00$~bohr proton-proton distance, as well as for the ground state of \Htwo{} molecule with $R=1.40$~bohr. In all cases our results agree to 5-6 significant digits with the most precise reference data (Table~\ref{tab:BLtest}). 

Table~\ref{tab:BL} collects original results computed in this work. 
There is currently no data available for the excited states of the hydrogen molecule. 
We have computed the Bethe logarithm for the three lowest electronically excited $^1\Sigma_\text{g}$ states, $EF$, $GK$, and $H\bar{H}$, and for the lowest two $^1\Sigma_\text{u}^+$ states, $B$ and $B'$, of \Htwo{} near the local minima of the Born--Oppenheimer potential energy wells (and for an intermediate value between the two minima of the $EF$ state). 

The expectation value of the $\delta(\br_{ia})$ operator was taken from Wolniewicz for the $EF$, $GK$, $H\bar{H}$ and $B$ states \cite{wolniewicz95,Wolniewicz98}, and for the $B'$ state it was evaluated using the integral-transformation technique in the present work \cite{PaCeKo05,JeIrFeMa22}.

The Bethe logarithm of the simplest polyatomic system, \Hthreeplus{} is evaluated near the equilibrium configuration with $D_{3\text{h}}$ symmetry. The precise value of $\langle\delta(\br_{ia})\rangle$ was taken from our earlier work, Ref.~\citenum{JeIrFeMa22}. We note that our $\blog$ value is close to the value given without any explanation or reference, $\ln k_0=3.0085$, in the \som\ of Ref.~\citenum{JaLe20} . 

To the best of our knowledge, the Bethe logarithm is evaluated for the first time for a point on a reactive potential energy surface. In the equilibrium configuration of the H$_3$ complex, the three hydrogen atoms are collinear \cite{FeMa22}. The computation was carried out near the equlibrium geometry with $R_{\text{\Htwo }}=1.4$~bohr and the distance of the third H atom from the center of mass of the \Htwo{} fragment was $R_{\mathrm{H}\cdots \mathrm{H}_2}=6.442$~bohr. 

The lowest rotational and rovibrational states of the \Hetwop{} molecular ion have been recently computed by taking into account the non-adiabatic, leading-order relativistic, and leading-order QED corrections \cite{FeKoMa2020}. The Bethe logarithm was estimated based on the observation that its value is only weakly dependent on the number of electrons, but strongly dependent on the nuclear charge. Based on this observation \cite{FeKoMa2020}, its value for He$_2^+$ (ground state) was approximated with the $\blog$ value of the one-electron He$_2^{3+}$ (ground state) ion,
$\blog(\text{He}_2^{3+})=4.388$ for $R=2.00$~bohr nuclear separation \cite{FeKoMa2020}. In the present work, the explicit computation for the three-electron He$_2^+$ provides a numerical check and confirmation of this approximation.

\vspace{0.5cm}
\begin{table}[h]
  \caption{%
    The Bethe logarithm computed in the present work for selected electronically excited states of the H$_2$ molecule, the ground state of H$_3^+$ and H$_2+$H at the equilibrium structures. 
    \label{tab:BL}}
 \begin{tabular}{@{}l lc l @{}}
    \hline\hline\\[-0.35cm]
    & & $R$ & \multicolumn{1}{c}{$\ln k_0$} \\
    \cline{1-4} \\[-0.3cm]
    H$_2$ 
    & $EF\ ^1\Sgp$  & $1.90$  & $ 3.000\ 8(4)$ \\
    &   & $3.00$  &  $2.985\ 8(5)$  \\
    &   & $4.40$  &  $2.978\ 0(1)  $  \\
    \cline{2-4} \\[-0.3cm]  
    &   $GK\ ^1\Sgp$  & $2.00$ & $3.008\ 7(5)$ \\
    &   & $3.20$  &     $2.992\ 7(1)$ \\
    \cline{2-4} \\[-0.3cm]  
    &   $H\bar{H}\ ^1\Sgp$  & $2.00$  &  $2.745\ 0(2)$ \\
    &   & $11.20$   & $  2.858\ 7(7)$
      \\
    \cline{2-4} \\[-0.3cm] & $B\ ^1\Sup$  & $2.40$ &
    $3.012\ 0(1)$  \\
    \cline{2-4} \\[-0.3cm] & $B'\ ^1\Sup$  & $2.00$  &
    $3.318\ 9(4)$ \\ 
    \cline{1-4} \\[-0.3cm]
    H$_3^+$ &	
    $1\ ^1A_1' $  & $1.65^\text{a}$  & $3.022(1) $ \\
    \cline{1-4} \\[-0.3cm]
    H$_2+$H &	
    $1\ ^2A_1 $  & $6.442^\text{b}$  &  $3.00(1)$   \\
    \cline{1-4} \\[-0.3cm]
    He$_2^+ $ &	
    $X\ ^2\Sup $  & $2.00$  &  $4.373(1)$   \\
    \hline\hline
  \end{tabular}
  \begin{flushleft}
    $^\text{a}$ %
      Equilateral triangular configuration with $R=1.65$~bohr. \\
    $^\text{b}$ %
      Collinear configuration, the proton-proton distance of H$_2$  is $R_{\mathrm{H}_2}=1.4 $ bohr and the distance of the H atom from the center of nuclear mass of the H$_2$ unit is $R_{\mathrm{H}\cdots \mathrm{H}_2}=6.442$ bohr.
  \end{flushleft}
\end{table}

\section{Summary and conclusion}
\noindent An efficient implementation of the non-relativistic Bethe logarithm
has been reported and summarized using a Hylleraas functional approach, numerically stabilized quadrature integration of the low- and intermediate-, and analytic integration of the asymptotic expansion for the high-photon momentum range.
The developed approach has been used to reproduce benchmark data to 5-6 significant digits (using double precision arithmetic),
and was newly employed to pinpoint the value of the Bethe logarithm for a series of small polyelectronic and polyatomic systems potentially relevant for precision spectroscopy, ultra-cold chemistry and reaction dynamics.

\section{Acknowledgement}
Financial support of the European Research Council through a Starting Grant (No.~851421) is gratefully acknowledged. DF thanks a doctoral scholarship from
the ÚNKP-21-3 New National Excellence Program of the Ministry for Innovation and Technology from the source of the National Research, Development and Innovation Fund (ÚNKP-21-3-II-ELTE-41).


\clearpage

\setcounter{section}{0}
\renewcommand{\thesection}{S\arabic{section}}
\setcounter{subsection}{0}
\renewcommand{\thesubsection}{S\arabic{section}.\arabic{subsection}}

\setcounter{equation}{0}
\renewcommand{\theequation}{S\arabic{equation}}

\setcounter{table}{0}
\renewcommand{\thetable}{S\arabic{table}}

\setcounter{figure}{0}
\renewcommand{\thefigure}{S\arabic{figure}}

~\\[0.cm]
\begin{center}
\begin{minipage}{0.8\linewidth}
\centering
\textbf{Supplementary Material} \\[0.25cm]

\textbf{Evaluation of the Bethe logarithm: from atom to chemical reaction}
\end{minipage}
~\\[0.5cm]
\begin{minipage}{0.6\linewidth}
\centering

D\'avid Ferenc,$^1$ and Edit M\'atyus$^{1,\ast}$ \\[0.15cm]

$^1$~\emph{ELTE, Eötvös Loránd University, Institute of Chemistry, 
Pázmány Péter sétány 1/A, Budapest, H-1117, Hungary} \\[0.15cm]
$^\ast$ edit.matyus@ttk.elte.hu \\
\end{minipage}
~\\[0.15cm]
(Dated: November 15, 2022)
\end{center}

~\\[1cm]
\begin{center}
\begin{minipage}{0.9\linewidth}
\noindent %
Contents: \\
S1. The Thomas--Reiche--Kuhn sum rule  \\
S2. Auxiliary quatities \\
\end{minipage}
\end{center}

\clearpage

\section{The Thomas--Reiche--Kuhn sum rule}
Let us consider a general $N$-particle Hamiltonian without external magnetic field with equal particle masses set to unity
\begin{align}
    H=\frac{1}{2}\sum_{i=1}^N \bp_i^2+V(\br_1,\dots,\br_N)
\end{align}
Define the total momentum and coordinate operators
\begin{align}
    \bos{P}=\sum_{i=1}^N\bp_i  && \bos{R} = \sum_{i=1}^N \br_i 
\end{align}
Their commutator is proportional to the number of particles $(N)$.
\begin{align}
    \left[\bos{P},\bos{R}\right]=-\iim N && \left[\grad,\bos{R}\right]  = N
\end{align}
Let us take the matrix element of the $\mu$th Cartesian component ($\mu=x,y,z$) of the above commutator with the $\ket{0}$ ground state eigenfunction of $H$ and insert the resolution of identity
\begin{align}
    &\matrixel{0}{\left[\grad_\mu ,\bos{R}_\mu \right]}{0} = \matrixel{0}{\grad_\mu \bos{R}_\mu}{0}-\matrixel{0}{\bos{R}_\mu\grad_\mu }{0} \\ \nonumber
    &=\sum_k \left[\matrixel{0}{\grad_\mu }{k} \matrixel{k}{ \bos{R}_\mu}{0} - \matrixel{0}{\bos{R}_\mu}{k} \matrixel{k}{\grad_\mu }{0} \right] \\ \nonumber
    &= \sum_k \left[\matrixel{0}{\grad_\mu }{k} \matrixel{k}{ \bos{R}_\mu}{0} + \matrixel{0}{\grad_\mu }{k} \matrixel{0}{\bos{R}_\mu}{k} \right] \\ \nonumber  
    &= 2\sum_k \matrixel{0}{\grad_\mu }{k} \matrixel{k}{ \bos{R}_\mu}{0} 
    \label{eq:TRk1}
\end{align}
Let us calculate the matrix elements of the commutator of $H$ and $\bos{R}_\mu$.
\begin{align}
    \matrixel{n}{\left[H,\bos{R}_\mu\right]}{k} = \matrixel{n}{H\bos{R}_\mu}{k}-\matrixel{n}{\bos{R}_\mu H}{k}=(E_n-E_k) \matrixel{n}{\bos{R}_\mu}{k}
\end{align}
Since only the kinetic term (it's $\mu$th component) does not commute with $\bos{R}_\mu$, we may write
\begin{align}
    &\matrixel{n}{\left[H,\bos{R}_\mu\right]}{k}
    =
    \frac{1}{2}\matrixel{n}{\left[\bos{P}^2_\mu,\bos{R}_\mu\right]}{k} \\ \nonumber
    &= 
    \frac{1}{2} \left[ \matrixel{n}{\left[\bos{P}_\mu,\bos{R}_\mu\right] \bos{P}_\mu }{k} 
    + 
    \matrixel{n}{ \bos{P}_\mu \left[\bos{P}_\mu.\bos{R}_\mu\right]}{k} \right] \\ \nonumber
    &=-\iim\matrixel{n}{\bos{P}_\mu}{k} = -\matrixel{n}{\grad_\mu}{k}
\end{align}
Thus, we arrive at the following relation between the momentum and coordinate matrix elements
\begin{align}
    \matrixel{n}{\bos{R}_\mu}{k} = -\frac{\matrixel{n}{\grad_\mu}{k}}{E_n-E_k} \; .
\end{align}
Summing up for the Cartesian components, dividing by three and plugging this back to Eq.~(\ref{eq:TRk1}) we get
\begin{align}
    N = -\frac{2}{3}\sum_k\frac{\abs{\matrixel{0}{\grad}{k}}^2}{E_k-E_0} \; ,
\end{align}
which gives the analytic value of $J(0)$ (see also pp. 256 of Ref.~\cite{BeSaBook57}).

\newpage
\section{Auxiliary quantities}
\begin{table}[h]
  \caption{%
    Auxiliary quantities used for the evaluation of the Bethe logarithm (Tables~II and III).
    The values taken from the literature are labelled with the corresponding reference. All other values are computed in this work. To have a precise value of $\langle\sum_{ia}\delta(\bos{r}_{ia})\rangle$,
    we used the integral transformation technique \cite{PaCeKo05} generalized for triatomics in Ref.~\cite{JeIrFeMa22}.
    $\kappa_1$ and $\kappa_2$ define the three photon momentum intervals (Sec.~IIA).
    \label{tab:BL}
    }
 \begin{tabular}{@{}l l@{\ \ } c @{\ \ } c @{\ \ \ } c @{\ \ \ } c @{\ \ \ } c @{\ \ \ }  c }
    \hline\hline\\[-0.35cm]
	&	State	&	$R$	&	$E_0$	&	$ \langle \sum_{ia} \delta (\br_{ia}) \rangle$	&	$\langle \grad^2 \rangle $	&	$\kappa_1$	&	$\kappa_2$	\\	\cline{1-8} \\[-0.3cm]
He	&	$1\ ^1S$	  &	\hspace{2cm}	&	$-$2.903 724 377	&	3.620 858 \cite{KorobovKorobov99}	&	$-$6.125 587 704	&	50	&	1000	\\	
	&	$2\ ^1S$	        &	    	&	$-$2.145 974 045	&	2.618 921 \cite{KorobovKorobov99}	&	$-$4.310 955 821	&	50	&	1000	\\	\cline{1-8} \\[-0.3cm]
H$_2^+$             &		&	2.00    &	$-$0.602 634 211	&	0.209 669 \cite{BuJeMoKo92}	&	$-$1.204 116 491	&	10	&	1000	\\	\cline{1-8} \\[-0.3cm]
H$_2$&  $X\ ^1\Sigma_g^+$	&	1.40	&	$-$1.174 475 714	&	0.459 668 \cite{PuKoPa17}	&	$-$2.551 772 866	&	10	&	1000	\\	
	&	$EF\ ^1\Sigma_g^+$	&	1.90	&	$-$0.718 147 440	&	0.232 764 \cite{Wolniewicz98}	&	$-$1.460 245 109	&	10	&	1000	\\	
	                &		&	3.00	&	$-$0.690 747 051	&	0.188 525 \cite{Wolniewicz98}	&	$-$1.298 146 037	&	10	&	1000	\\	
	                &		&	4.40	&	$-$0.714 520 545	&	0.200 146 \cite{Wolniewicz98}	&	$-$1.403 027 424	&	10	&	1000	\\	
	&	$GK\ ^1\Sigma_g^+$	&	2.00	&	$-$0.660 429 778	&	0.209 591 \cite{Wolniewicz98}	&	$-$1.317 256 686	&	10	&	1000	\\	
	&	                	&	3.20	&	$-$0.662 692 409	&	0.185 240 \cite{Wolniewicz98}	&	$-$1.224 961 538	&	10	&	1000	\\	
& $H\bar{H}\ ^1\Sigma_g^+$	&	2.00	&	$-$0.654 696 286	&	0.213 700 \cite{Wolniewicz98}	&	$-$1.346 869 874	&	10	&	1000	\\	
    &               		&	11.20	&	$-$0.605 302 323	&	0.168 760 \cite{Wolniewicz98}	&	$-$1.214 522 002	&	5	&	300	    \\	
	&	$B\ ^1\Sigma_u^+$	&	2.40	&	$-$0.756 674 940	&	0.206 265 \cite{wolniewicz95}	&	$-$1.846 078 900	&	10	&	1000	\\	
	&	$B'\ ^1\Sigma_u^+$	&	2.00	&	$-$0.665 481 264	&	0.227 450 \cite{wolniewicz95}	&	$-$1.421 340 210	&	10	&	1000	\\	\cline{1-8} \\[-0.3cm]
H$_3^+$	&	$1\ A_1'$	    &	1.65	&	$-$1.343 835 625	&	0.363 218 \cite{JeIrFeMa22}	&	$-$2.924 659 015	&	10	&	1000	\\	\cline{1-8} \\[-0.3cm]
H$_3$	&	$1\ A_1$	    &	6.442$^\text{a}$	&	$-$1.674 561 687	&	0.412 498	&	$-$3.549 629 723	&	10	&	1000	\\	\cline{1-8} \\[-0.3cm]
He$_2^+$& $X\ ^2\Sigma_u^+$	&	2.00	&	$-$4.994 441 731	&	3.144 192	&	$-$10.201 626 980	&	10	&	1000	\\	
    \hline\hline
  \end{tabular}
  \begin{flushleft}
         $^\text{a}$ %
         Linear configuration, the H$_2$ bond length is 1.40 bohr. Here $R$ is the distance of the H atom from the center of mass of the H$_2$ unit. 
      \end{flushleft}
\end{table}

\clearpage
\bibliography{references}

\end{document}